\documentclass[reprint,superscriptaddress,amsmath,amssymb,aps,prl,floatfix]{revtex4-2}
\usepackage{pdfpages,pgffor,graphicx,dcolumn,bm,xr,physics,chemformula,siunitx,color,soul,float}

\newcommand{\bmk}{{\bf k}}
\newcommand{\bmq}{{\bf q}}
\newcommand{\bmqq}{{\bf Q}}
\newcommand{\bmr}{{\bf r}}
\newcommand{\bmrr}{{\bf R}}

\newcommand{\ifc}{C_{\kappa \alpha p, \kappa' \alpha' p'} }
\newcommand{\iifc}{C^{-1}_{\kappa \alpha p, \kappa' \alpha' p'} }

\newcommand{\dtau}{\Delta \tau}

\newcommand{\kpa}{{\kappa \alpha p }}
\newcommand{\kpaa}{{\kappa' \alpha' p' }}

\newcommand{\vck}{{vc\bmk }}

\newcommand{\fourr}{{ \bmr_{\rm e}, \bmr_{\rm h}; \bmr_{\rm e}', \bmr_{\rm h}'}}

\newcommand{\dfourr}{d\bmr_{\rm e} d\bmr_{\rm h} d\bmr_{\rm e}' d\bmr_{\rm h}'}

\makeatletter
\AtBeginDocument{\let\LS@rot\@undefined}
\makeatother

\colorlet{blue}{blue!70!black} 

\colorlet{red}{red!70!black}

\begin{document}

\title{Excitonic polarons and self-trapped excitons \\[3pt] from first-principles exciton-phonon couplings}

\author{Zhenbang Dai}
\affiliation{Oden Institute for Computational Engineering and Sciences, The University of Texas at Austin, Austin, Texas 78712, USA}
\affiliation{Department of Physics, The University of Texas at Austin, Austin, Texas 78712, USA}
\author{Chao Lian}
\affiliation{Oden Institute for Computational Engineering and Sciences, The University of Texas at Austin, Austin, Texas 78712, USA}
\affiliation{Department of Physics, The University of Texas at Austin, Austin, Texas 78712, USA}
\author{Jon Lafuente-Bartolome}
\affiliation{Oden Institute for Computational Engineering and Sciences, The University of Texas at Austin, Austin, Texas 78712, USA}
\affiliation{Department of Physics, The University of Texas at Austin, Austin, Texas 78712, USA}
\author{Feliciano Giustino}%
\email{fgiustino@oden.utexas.edu}
\affiliation{Oden Institute for Computational Engineering and Sciences, The University of Texas at Austin, Austin, Texas 78712, USA}
\affiliation{Department of Physics, The University of Texas at Austin, Austin, Texas 78712, USA}
        
\date{\today}

\begin{abstract}
Excitons consist of electrons and holes held together by their attractive Coulomb interaction. Although excitons are neutral excitations, spatial fluctuations in their charge density couple with the ions of the crystal lattice. This coupling can lower the exciton energy and lead to the formation of a localized excitonic polaron, or even a self-trapped exciton in the presence of strong exciton-phonon interactions. Here, we develop a theoretical and computational approach to compute excitonic polarons and self-trapped excitons from first principles. Our methodology combines the many-body Bethe-Salpeter approach with density-functional perturbation theory, and does not require explicit supercell calculations. As a proof of concept, we demonstrate our method for a compound of the halide perovskite family.
\end{abstract}

\maketitle

One of the most striking manifestation of electron-phonon interactions in solids is the formation of polarons. These quasiparticles form when an electron creates a small distortion of the surrounding crystal lattice, which in turn acts as a potential well that promotes electron localization~\cite{franchini2021polarons}. Polarons have been investigated extensively in a wide variety of materials, including alkali halides~\cite{sio2019polarons,iadonisi1984electron,inoue1970electronic, lee2021facile}, metal oxides~\cite{setvin2014direct, yang2013intrinsic, frederikse1964electronic,falletta2022polarons}, hybrid perovskites~\cite{guzelturk2021visualization, mayers2018lattice, miyata2017large, zhu2015charge}, and two-dimensional materials~\cite{sio2023polarons}.  
% {\color{red}[we should cite Bernardi and Pasquarello]} 

A similar feedback mechanism between charge and lattice can occur for neutral excitations such as excitons~\cite{rohlfing1998electron,rohlfing2000electron}.
% {\color{red}[cite Louie-Rohlfing]}
While the exciton is charge-neutral overall, local spatial fluctuations of its electron and hole charge densities can induce lattice distortions as in the case of polarons. In turn, these distortions can stabilize the exciton and promote its localization. The resulting quasiparticle is referred to as an excitonic polaron~\cite{iadonisi1987polaronic}.
% {\color{red}[textbook?]}. 
In the presence of strong exciton-phonon interactions, the same mechanism can lead to the formation of self-trapped excitons~\cite{fowler1973theory, ismail2005self, williams1990self, li2020self}
\footnote{{We note that the names ``excitonic polarons'' and ``self-trapped excitons'' refer to the same physical mechanism of polaronic stabilization of the exciton. The difference between these two concepts resides in the strength of the electron-phonon coupling, the resulting phonon-induced localization of the exciton wavefunction, and the magnitude of the hopping barrier for exciton migration. Therefore, in this Letter, we use the two naming conventions interchangeably.}}.

Excitonic polarons are often invoked to explain large Stokes shifts observed between absorption and luminescence spectra~\cite{ismail2005self, abfalterer2020colloidal, williams1990self}. 
In these cases, the photoluminescence red-shift is interpreted as the result of the lattice relaxation in the optically-excited state, within a Franck-Condon picture~\cite{nitzan2006chemical}.
% {\color{red}[textbook?]}. 
These conceptual models are ubiquitous in solid-state spectroscopy and have been in use for decades, but detailed \textit{ab initio} calculations remain exceedingly rare~\cite{mauri1995first, ismail2003excited, sattelmeyer2001comparison,stanton1995comparison}. The main difficulties are that (i) calculations of Hellman-Feynman forces in the excited state using the Bethe-Salpeter equation (BSE) approach are challenging~\cite{ismail2003excited}; (ii) since exciton localization breaks the translational invariance of the crystal unit cell, BSE calculations on large supercells are needed~\cite{ismail2005self}, and this poses a significant computational challenge.
%For example, the authors of Ref.~\citenum{ismail2005self} used a supercell to investigate the self-trapped exciton in \ch{SiO2}, but the need for solving the BSE in a supercell limits the applicability of this approach.

Here, we develop an \textit{ab initio} theoretical and computational method for calculating excitonic polarons and self-trapped excitons without resorting to supercells. 
{Following a strategy similar in spirit to previous work} on polarons~\cite{sio2019polarons, sio2019ab, sio2023polarons, lafuente2022ab, lafuente2022unified}, we express the wavefunction of the excitonic polaron as a coherent superposition of finite-momentum excitons, and we recast the BSE total energy functional in the excited state as a self-consistent eigenvalue problem in the exciton coefficients. Alongside the excitonic wavefunction, our present approach provides the accompanying atomic displacements and their spectral decomposition into normal vibrational modes, thereby offering a detailed picture of the excitonic polaron {and the atomic-scale mechanisms that drive its formation}. To demonstrate this method, we investigate {excitonic polarons} in \ch{Cs2ZrBr6}, a vacancy-ordered double perovskite {that attracted interest in the context of energy-efficient lighting, and which exhibits signatures of exciton self-trapping}~\cite{abfalterer2020colloidal, cucco2022fine}. Additional \textit{ab initio} calculations of excitonic polarons in LiF, applications to model Hamiltonians, and a more in-depth analysis of the theoretical formalism are presented in the companion manuscript~\cite{dai2023explrn_prb}.

The total energy of a crystal in a neutral excited state can be expressed as~\cite{ismail2003excited}:
\begin{eqnarray}
    \label{eqn:etot_rs}
    &&E\left[ \Psi(\bmr_{\rm e},\bmr_{\rm h}) ,
    \{ \Delta \tau_{\kpa} \}
    \right]
    = E_0\nonumber \\
    &&
    \qquad +\int_{\rm sc}
    \Psi^*(\bmr_{\rm e},\bmr_{\rm h}) H_{\mathrm{BSE}}(\fourr) \Psi(\bmr_{\rm e}',\bmr_{\rm h}')
    d\bmr
    \nonumber \\
    &&\hspace{20pt}+
    \,\frac{1}{2}\!\! \sum_{\substack{\kpa\\ \kpaa}} \ifc \Delta \tau_{\kpa} \Delta \tau_{\kpaa},
\end{eqnarray}
where $E_0$ denotes the ground-state energy with the atoms in their equilibrium positions;
$H_{\mathrm{BSE}}$ is the BSE Hamiltonian~\cite{rohlfing1998electron, rohlfing2000electron}; $\Psi(\bmr_{\rm e},\bmr_{\rm h})$ is the exciton wavefunction, with 
$\bmr_{\rm e}$ and $\bmr_{\rm h}$ denoting the electron and hole coordinates, respectively. The integral extends over the Born-von K\'arm\'an (BvK) supercell, and the integration variable $d\bmr$ is a short-hand notation for $\dfourr$. $\ifc$ denotes the matrix of interatomic force constants, and $\Delta \tau_{\kpa}$ is the displacement of the atom $\kappa$ in the unit cell $p$ along the Cartesian direction $\alpha$, with respect to the ground-state structure. The term on the last line of Eq.~(\ref{eqn:etot_rs}) describes the elastic energy associated with the lattice distortion. {The energy} functional is truncated to the second order in the displacements; this approximation proved successful in calculations of both small and large polarons, comparing well with direct hybrid-functional calculations~\cite{sio2019polarons, sio2019ab, sio2023polarons, lafuente2022ab, lafuente2022unified}.
In Eq.~\eqref{eqn:etot_rs}, the BSE Hamiltonian and wavefunctions depend implicitly on the atomic positions, and the interatomic force constants are assumed to be the same for the ground and excited state. This approximation was shown to be reliable for large and small polarons \cite{sio2019ab}.
%, and could be lifted for more refined calculations.

\raggedbottom

To obtain the exciton wavefunction and the atomic displacements that minimize the total energy in Eq.~\eqref{eqn:etot_rs}, we use the method of Lagrange multipliers, and set to zero the functional derivatives of $E\left[ \Psi(\bmr_{\rm e},\bmr_{\rm h}) , \{ \Delta \tau_{\kpa} \} \right]$ with respect to $\Psi(\bmr_{\rm e},\bmr_{\rm h})$ and $\{ \Delta \tau_{\kpa} \}$, subject to the normalization constraint $\int_{\rm sc}\abs{\Psi(\bmr_{\rm e},\bmr_{\rm h})}^2 d\bmr_{\rm e} d\bmr_{\rm h}=1$. After some algebra, we arrive at the following coupled nonlinear eigenvalue problem:
\begin{eqnarray}
    \label{eqn:minimize_psi}
    &&\int_{\rm sc}\!\!
    H^0_{\mathrm{BSE}}(\fourr) \,\Psi(\bmr_{\rm e}',\bmr_{\rm h}')\,d\bmr_{\rm e}' d\bmr_{\rm h}' 
    \nonumber \\
    &+&
    \sum_{\kpa}
    \int_{\rm sc}
    \frac{\partial H^0_{BSE}(\fourr)}{\partial \tau_{\kpa}}\,
    \Psi(\bmr_{\rm e}',\bmr_{\rm h}')\,
    d\bmr_{\rm e}' d\bmr_{\rm h}'\, \dtau_{\kpa}
    \nonumber \\
    &=& \varepsilon \,\Psi(\bmr_{\rm e},\bmr_{\rm h}), \\
    \label{eqn:minimize_tau}
    &&\hspace{-15pt}\dtau_{\kpa} 
    = 
    -\!\!\sum_{\kappa'\alpha' p'}\iifc
    \nonumber \\
    &&\times\int_{sc} \!\!
    \Psi^*(\bmr_{\rm e},\bmr_{\rm h})  
    \frac{\partial H^0_{BSE}(\fourr)}{\partial \tau_{\kpaa}}
    \Psi(\bmr_{\rm e}',\bmr_{\rm h}')\,d\bmr,
\end{eqnarray}
where $H^0_{\mathrm{BSE}}$ is the BSE Hamiltonian for the undistorted system, and the eigenvalue $\varepsilon$ is the Lagrange multiplier. 

To solve Eqs.~(\ref{eqn:minimize_psi}) and (\ref{eqn:minimize_tau}) without resorting to supercells, we express $\Psi(\bmr_{\rm e},\bmr_{\rm h})$ {\color{black}as} a linear combination of exciton states $\Omega_{s\bmqq}(\bmr_{\rm e},\bmr_{\rm h})$ of the undistorted ground-state structure, and we express the displacements $\Delta \tau_{\kpa}$ as a linear combination of normal vibrational modes of the undistorted structure:
\begin{eqnarray}
    \label{eqn:exciton_basis}
    &&\Psi(\bmr_{\rm e},\bmr_{\rm h})
    =
    \frac{1}{\sqrt{N}}
    \sum_{s\bmqq} A_{s\bmqq}\, \Omega_{s\bmqq}(\bmr_{\rm e},\bmr_{\rm h}), \\
        \label{eqn:disp_pattern}
    &&\Delta \tau_{\kpa}
    =
    -\frac{2}{N} \sum_{\bmq \nu}
    B_{\bmq \nu} \left(\frac{\hbar}{2M_{\kappa}\omega_{\bmq \nu}}\right)^{1/2}
    e_{\kappa \alpha, \nu}(\bmq) e^{i\bmq\cdot\bmrr_p}. \nonumber \\
\end{eqnarray}
In these expressions, $s$ is the exciton band index, $\bmqq$ is the exciton momentum, and {\color{black}$N$} is the number of unit cells in the BvK supercell; $M_{\kappa}$ is the mass of atom $\kappa$, $e_{\kappa \alpha, \nu}(\bmq)$ is polarization vector of the phonon with momentum $\bmq$, branch $\nu$, and frequency $\omega_{\bmq\nu}$; $\bmrr_p$ is the lattice vector of the $p$-th unit cell in the BvK supercell~\cite{giustino2017electron}. The sums are carried out over uniform Brillouin zone grids with {\color{black}$N$} points. The expansion coefficients $A_{s\bmqq}$ and $B_{\bmq \nu}$ in Eqs.~(\ref{eqn:exciton_basis}) and (\ref{eqn:disp_pattern}) can be interpreted as the contribution of each exciton and each phonon of the undistorted lattice to the excitonic polaron, respectively.

{\color{black}The exciton states of the undistorted lattice appearing in Eq.~\eqref{eqn:exciton_basis}, $\Omega_{s\bmqq}(\bmr_{\rm e},\bmr_{\rm h})$, can be expressed} in terms of Kohn-Sham single-particle states. To this end, we employ the Tamm-Dancoff approximation~\cite{onida2002electronic, rohlfing2000electron}:
\begin{align}
    \label{eqn:tda}
    \Omega_{s\bmqq}(\bmr_{\rm e},\bmr_{\rm h}) 
    =
    \sum_{\vck} a_{\vck}^{s\bmqq} 
    \psi_{v\bmk}^*(\bmr_{\rm h}) \psi_{c\bmk+\bmqq}(\bmr_{\rm e}),
\end{align}
where $\psi_{n\bmk}(\bmr)$ are Kohn-Sham eigenstates of the undistorted ground-state, $v$ and $c$ refer to valence and conduction bands, respectively, and the coefficients $a^{s\bmqq}_{\vck}$ denote the  eigenvectors of the ground-state BSE Hamiltonian, $H^0_\mathrm{BSE}$, with eigenvalue $E^0_{s\bmqq}$.

% {\color{red}[avoiding braket notation for consistency: $\braket{s\bmqq | \mathcal{G}^{\nu} | s'\bmqq'}$ becomes $\mathcal{G}_{ss'\nu}(\bmqq,\bmqq')$]}
Upon substituting Eqs.(\ref{eqn:exciton_basis})-(\ref{eqn:tda}) inside Eqs.~(\ref{eqn:minimize_psi})-(\ref{eqn:minimize_tau}), we obtain a coupled nonlinear system of equations for the coefficients $A_{s\bmqq}$ and $B_{\bmq \nu}$:
\begin{eqnarray}
    \label{eqn:explrneqn}
    &&\sum_{s'\bmqq'} 
    \bigg[ 
    E^0_{s\bmqq}
    \delta_{ss'}
    \delta_{\bm{Q} \bm{Q}'} 
    -\frac{2}{N}\sum_{\nu}
    B_{\bmqq-\bmqq' \nu} \mathcal{G}_{ss'\nu}(\bmqq',\bmqq-\bmqq')
    \bigg] 
    \nonumber \\ && \hspace{20pt}\times A_{s'\bmqq'}
    = \varepsilon A_{s\bmqq},  \\
    &&B_{\bmqq \nu}  
    = 
    \frac{1}{N \hbar \omega_{\bmqq \nu}} 
    \sum_{\substack{ss' \bmqq'}}
    A_{s'\bmqq'}^* A_{s\bmqq'+\bmqq}
 \mathcal{G}^*_{ss'\nu}( \bmqq', \bmqq), \nonumber \\[-10pt] \label{eqn:bmat}
\end{eqnarray}
having introduced the exciton-phonon {\color{black}coupling} matrix element~\cite{chen2020exciton, antonius2022theory}:
\begin{align}
    \label{eqn:exphg}
    \mathcal{G}_{ss'\nu}(\bmqq,\bmq)
    =&
    \sum_{\vck} a_{\vck}^{s\bmqq+\bmq*}  
    \Bigg[
    \sum_{c'}
    g_{cc'\nu} (\bmk+\bmqq,\bmq)
    a_{vc'\bmk}^{s'\bmqq}
    \nonumber \\
    &-\sum_{v'}
    g_{v'v\nu} (\bmk,\bmq) 
    a_{v'c\bmk + \bmq}^{s'\bmqq}
    \Bigg].
\end{align}
In this {\color{black}expression}, $g_{mn\nu}(\bmk,\bmq)$ is the standard electron-phonon coupling matrix element between the Kohn-Sham states $n\bmk$ and $m\bmk+\bmq$ via the phonon $\bmq\nu$~\cite{giustino2017electron}. 

\raggedbottom

Equations~(\ref{eqn:explrneqn}) and (\ref{eqn:bmat}) constitute the central result of this work. The solution of these equations describes the formation of excitonic polarons and self-trapped excitons via the exciton-phonon interaction. 
% {\color{black}Equations~(\ref{eqn:explrneqn})-(\ref{eqn:bmat}) share the same formal structure as} the \textit{ab initio} polaron equations given in Refs.~\citenum{sio2019ab, sio2019polarons}; in fact, by replacing the BSE Hamiltonian with the Kohn-Sham Hamiltonian, and by replacing the exciton-phonon coupling matrix elements $\mathcal{G}_{ss'\nu}(\bmqq,\bmq)$ with the electron-phonon coupling matrix elements $g_{mn\nu}(\bmk,\bmq)$, {\color{black}one recovers} the polaron equations of Ref.~\citenum{sio2019ab}.
All quantities appearing in Eqs.~(\ref{eqn:explrneqn})-(\ref{eqn:exphg}) can be obtained from calculations performed in the crystal unit cell, without requiring supercells. The breaking of lattice periodicity comes from the coherent superposition of exciton states with different momenta. The largest size of the exciton polaron that can be studied is set by the BvK supercell, which in the present formalism translates into the number $N$ of grid-points in the Brillouin zone. 

% {\color{red}[pls cross-check/correct this sentence, I probably used improper language]} 
% \red{[Z: removed discussion of eigenvalue since Fig. 3c has been removed]}
%{The eigenvalue $\varepsilon$ in Eq.~\eqref{eqn:explrneqn} is such that $E_{\rm gap}^{\rm GW}\!-\varepsilon$ represents the vertical dissociation energy of the excitonic polaron at fixed atomic positions, as shown schematically in Fig.~\ref{fig:energies}(c) ($E_{\rm gap}^{\rm GW}$ is the GW quasiparticle gap).
The dissociation energy of the excitonic polaron {into fully delocalized free electron-hole pair} is $E_{\rm gap}^{\rm GW}\!-E_{\rm xp}$, where $E_{\rm xp}$ is given by: 
% {\color{red}[Z: can we rewrite this equation in a form more similar to Eq. 41 of Sio2019PRB?]}
% \begin{eqnarray}
%     \label{eqn:etot_plrn}
%     E_{\rm xp} &=& \frac{1}{N_p}
%     \sum_{s\bmqq}
%     |A_{s\bmqq}|^2
%     E^0_{s\bmqq} \nonumber \\
%   &-& \frac{1}{N_p^2}
%     \sum_{\substack{ss'\nu\\ \bmqq \bmqq'}}
%     A_{s\bmqq}^*A_{s'\bmqq'}
%     B_{\bmqq-\bmqq' \nu} 
%     \mathcal{G}_{ss'\nu}(\bmqq,\bmqq').
% \end{eqnarray}
\begin{eqnarray}
    \label{eqn:etot_plrn}
    E_{\rm xp} = \frac{1}{N}
    \sum_{s\bmqq}
    |A_{s\bmqq}|^2
    E^0_{s\bmqq} 
   -\frac{1}{N}
    \sum_{\substack{\bmq \nu\\}}
    \abs{B_{\bmq\nu}}^2
    \hbar \omega_{\bmq \nu}
    .
\end{eqnarray}
The first term on the right-hand side of this equation represents the average energy of the exciton states participating to the excitonic polarons; the second term is the stabilization energy resulting from the lattice distortion. Similarly, the formation
energy of the excitonic polaron is $\Delta E_{\rm f}^{\rm xp}=E_{\rm xp} - E^0_{\rm ex}$, where $E^0_{\rm ex}$ is the energy of the lowest-lying exciton. 
% {\color{black}[removed repeated statement about the eigenvalue]}
%; {while the vertical (fixed lattice) formation energy is given by the eigenvalue $\varepsilon$ of Eq.~Eq.~\eqref{eqn:explrneqn}.}

To demonstrate the present methodology, we consider the vacancy-ordered halide double perovskite \ch{Cs2ZrBr6} as a test system. 
All calculations details are described in the Supplemental Information~\cite{SI}.
\ch{Cs2ZrBr6} crystallizes in the Fm$\bar 3$m space group, and consists of of a rock-salt lattice of alternating (ZrB$_6$)$^{2-}$ octahedra and vacancies, with Cs$^{1+}$ cations acting as spacers, see supplemental Fig.~S1. Optical absorption and photoluminescence spectra of this perovskite reveal a large Stokes shift, % of 1.78~eV, 
which has tentatively been assigned to the presence of self-trapped excitons~\cite{abfalterer2020colloidal, cucco2022fine}. 

In agreement with the experimental proposal, we do find a highly localized excitonic polaron in \ch{Cs2ZrBr6}.
Figure~\ref{fig:charge_density} shows the charge densities of the electron polaron, the hole polaron, and the excitonic polaron in \ch{Cs2ZrBr6}. We find small electron and hole polarons; in particular, the electron polaron in Fig.~\ref{fig:charge_density}(a) primarily consists of a single Zr-4$d$ orbital of $t_{\rm 2g}$ character ($d_{xz}$ and $d_{yz}$), while the hole polaron in Fig.~\ref{fig:charge_density}(b) is derived from a single Br-4$p$ orbital. Our findings are consistent with the Zr-$d$ and Br-$p$ characters of the conduction and valence band edges, respectively (supplemental Fig.~S2). The highly localized nature of polarons reflects the weakly-interacting nature of \ch{ZrBr6} octahedra in the vacancy-ordered double perovskite structure~\cite{cucco2022fine}.
In Figs.~\ref{fig:charge_density}(c) and (d) we show the electron density and the hole density in the excitonic polaron state, respectively, which are given by $n_{\rm e}(\bmr_{\rm e})=\int_{\rm sc}\abs{\Psi(\bmr_{\rm e},\bmr_{\rm h})}^2 d\bmr_{\rm h}$ and $n_{\rm h}(\bmr_{\rm h})=\int_{\rm sc}\abs{\Psi(\bmr_{\rm e},\bmr_{\rm h})}^2 d\bmr_{\rm e}$. As for the polarons, the charge densities of the excitonic polaron are localized in real space. By comparing Figs.~\ref{fig:charge_density}(b) and (d), we see that the hole density is distributed among all six Br atoms of a \ch{ZrBr6} octahedron, unlike the hole polaron which is localized on a single atom.

Localization is a distinctive feature of the excitonic polaron, whereas free excitons in the undistorted structure are fully delocalized. In fact, using Eq.~\eqref{eqn:tda} and the orthogonality of Kohn-Sham states, it is immediate to see that the charge densities of the excitons in the ground-state structure, $\int_{\rm sc}\abs{\Omega_{s\bmqq}(\bmr_{\rm e},\bmr_{\rm h})}^2 d\bmr_{\rm h}$ and $\int_{\rm sc}\abs{\Omega_{s\bmqq}(\bmr_{\rm e},\bmr_{\rm h})}^2 d\bmr_{\rm e}$, are lattice-periodic. This observation is consistent with the fact that, to plot the exciton wavefunction, one must fix the position of the hole to obtain a localized electron charge distribution~\cite{deslippe2012berkeleygw}. This type of plot is shown in Figs.~\ref{fig:charge_density}(e) and (f) for the excitons in \ch{Cs2ZrBr6}. 
% {\color{red}[removed sentenced]}
%These two states correspond to a dark and bright exciton, respectively, as we discuss later. 
% {\color{red} {By comparing Fig.~\ref{fig:charge_density}(c) with (e) and (f), we see that the excitonic polaron is more localized than the electron density of the exciton in the ground-state undistorted structure. This stronger localization is the hallmark of polaronic stabilization of free excitons.}}

\raggedbottom

Figure~\ref{fig:phonon} and supplemental Fig.~S3 show the atomic displacements accompanying the electron polaron, the hole polaron, and the excitonic polaron in \ch{Cs2ZrBr6}. 
The electron polaron, which is localized on the Zr atom, tends to attract the eight positively-charged nearest-neighbor Cs cations in a symmetric pattern [Fig.~S3(a)]. Conversely, the hole polaron, which is centered on a Br atom, tends to repel a single positively-charged nearest-neighbor Zr cation [Fig.~S3(b)]. 
The atomic displacement pattern of the excitonic polaron, shown in Fig.~\ref{fig:phonon}(a), 
resembles the pattern for the electron polaron in Fig.~S3(a), but additionally includes significant displacements of the nearest-neighbor Br atoms surrounding the central Zr atom.
% closely follows the pattern for the electron polaron in Fig.~\ref{fig:phonon}(a). 
This behavior can be rationalized by noting that, 
{due to the crystal symmetry of \ch{Cs2ZrBr6}, the hole polaron shown Fig.~\ref{fig:charge_density}(b) and Fig.~S3(b) is only one of the six degenerate solutions localized at the six Br atoms, respectively, and when forming the excitonic polaron, the six degenerate solutions contribute to making the hole density of the excitonic polaron carry the approximate spherical symmetry around the Zr atom. }
% in the excitonic polaron, the electronic density is concentrated around the central Zr cation, as in the electron polaron; at the same time, the hole density is distributed across six nearest-neighbor Br anions, unlike in the hole polaron. Therefore, the resulting charge density of the excitonic polaron carries approximate spherical symmetry around the Zr atom, as for the electron polaron. 

Our present formalism allows us to identify the phonon modes that are responsible for the formation of the excitonic polaron. Figure~\ref{fig:phonon}(b) shows the magnitude of the coefficients $|B_{\bmq\nu}|^2$ overlaid on the phonon dispersion relations of \ch{Cs2ZrBr6}. The dominant contributions arise from the phonon branches associated with the $A_{\rm 1g}$ mode at 23.2~meV and the $T_{\rm 1u}$ mode at 6.2~meV. The former is the octahedral breathing mode {\color{black}[Fig.~\ref{fig:phonon}(c)]}, and the latter corresponds to displacements of the Cs atoms {\color{black}[Fig.~\ref{fig:phonon}(d)]}. 
% Other modes shown in Fig.~\ref{fig:phonon}(b) contribute less than 8\% to the formation energy of the excitonic polaron. 
This decomposition is consistent with the atomic displacement patterns shown in Fig.~\ref{fig:phonon}(a). 
We emphasize that, while the largest atomic displacements occur for the Br atoms that are nearest neighbors to the central Zr atom, we find significant atomic displacements of Cs atoms in more distant neighbors within other unit cells. This result underscores the importance of being able to perform this type of calculations on large supercells, such as the 576-atoms equivalent supercell that we can access in the present study by solving Eqs.~(\ref{eqn:explrneqn})-(\ref{eqn:bmat}). {\color{black}Performing} direct BSE calculations on such large unit cells would be prohibitive using existing methods~\cite{mauri1995first, ismail2003excited, sattelmeyer2001comparison,stanton1995comparison}.

We now compare the formation energies for electron, hole, and excitonic polarons, as shown in Fig.~\ref{fig:energies}. 
{\color{black}Fig.~\ref{fig:energies}(a) shows a schematic illustration of the many-body quasiparticle energy diagram; Fig.~\ref{fig:energies}(b) shows the corresponding many-body BSE diagram for neutral excitations {\color{black}(exciton band structures and the BSE absorption coefficients are shown in supplemental Fig.~S4 for completeness)}.
We see that the} electron and hole polaron have relatively large formation energies ({\color{black}302~meV and 550~meV}, respectively) which can be rationalized in terms of the heavy electron and hole effective masses~\cite{sio2019ab, cucco2022fine}; instead, the formation energy of the excitonic polaron is of one magnitude smaller ({\color{black}48~meV}). 
{\color{black}In line with this weaker formation energy, we observe that the charge densities of the excitonic polaron [Figs.~\ref{fig:charge_density}(c) and \ref{fig:charge_density}(d)] are more diffuse than those of the electron polaron and of the hole polaron [Figs.~\ref{fig:charge_density}(a) and \ref{fig:charge_density}(b)].} 

{\color{black}The difference between the formation energy and localization of the polaron and of the excitonic polaron} can be {\color{black}rationalized by inspecting} the exciton-phonon coupling matrix element {\color{black}in} Eq.~\eqref{eqn:exphg}. {\color{black}In this expression,} the two terms associated with electron-phonon and hole-phonon couplings appear with opposite signs and tend to cancel out.
Physically, this {\color{black}partial cancellation stems from the fact that,} when small polarons with similar localization length and opposite charge combine to produce an exciton, the net charge density tends to cancel out, {which can be seen by the similar shape and distribution of the electron density and hole density of the excitonic polaron [Figs. ~\ref{fig:charge_density}(c) and \ref{fig:charge_density}(d)],} thus weakening the lattice distortion and leading to a smaller polaronic stabilization of the free exciton. 
This observation indicates that, to observe large polaronic effects in excitons, one needs to look for materials with large differences in the electron-phonon and hole-phonon interaction strengths. 

% \red{[Z: removed the discussion of the ZPL, which is a little out of context here just before the conclusions]}
%{\color{red} We also make quantitative comparison with experimental measurements on \ch{Cs2ZrBr6}~\cite{abfalterer2020colloidal}, as can be found in the Supplemental Information.  The good agreement between the theoretical and experimental zero-phonon line (Fig.~S5) supports the notion that the large Stokes shift observed in \ch{Cs2ZrBr6} is to be attributed to the formation of excitonic polarons.}

In summary, we developed a theoretical and computational method that enables, for the first time, \textit{ab initio} calculations of excitonic polarons and self-trapped excitons {\color{black}in real materials}. This method does not require computationally prohibitive supercell calculations, and can seamslessly be used to investigate small and large excitonic polarons as well as self-trapped excitons.
% The present method provides information on the dissociation energy of the excitonic polaron, on the wavefunction and atomic displacements, and allows one to unambiguously identify the lattice vibrations that drive the formation of the excitonic polaron state. In a first application to a light-emitting double perovskite, we we identified the formation of localized excitonic polarons, in agreement with experiments.
%The present development opens a number of interesting avenues, from the possibility of building simplified models for high-throughput studies to generalizations for including non-adiabatic and quantum nuclear effects. 
This work will make it possible to investigate the physics of exciton-phonon couplings and self-trapped excitons in diverse classes of materials with potential for solar energy harvesting, photocatalysis, energy-efficient lighting, and light-driven quantum matter.

\begin{acknowledgments}
We are grateful to Bruno Cucco and George Volonakis for bringing the self-trapped exciton in \ch{Cs2ZrBr6} to our attention. This research is primarily supported by the National Science Foundation, Office of Advanced Cyberinfrastructure and Division of Materials Research under Grant No. 2103991 of the Cyberinfrastructure for Sustained Scientific Innovation program, and the NSF Characteristic Science Applications for the Leadership Class Computing Facility program under Grant No. 2139536 (development of the excitonic polaron module, calculations, interoperability). This research was also supported by the Computational Materials Sciences Program funded by the US Department of Energy, Office of Science, Basic Energy Sciences, under award no. DE-SC0020129 (development of the polaron module). This research used resources of the National Energy Research Scientific Computing Center and the Argonne Leadership Computing Facility, which are DOE Office of Science User Facilities supported by the Office of Science of the U.S. Department of Energy, under Contracts No. DE-AC02-05CH11231 and DE-AC02-06CH11357, respectively. The authors acknowledge the Texas Advanced Computing Center (TACC) at The University of Texas at Austin for providing access to Frontera, Lonestar6, and Texascale Days, that have contributed to the research results reported within this paper (http://www.tacc.utexas.edu).
\end{acknowledgments}

\bibliography{literature}

% \clearpage

\begin{figure*}
\includegraphics[width=0.8\textwidth]{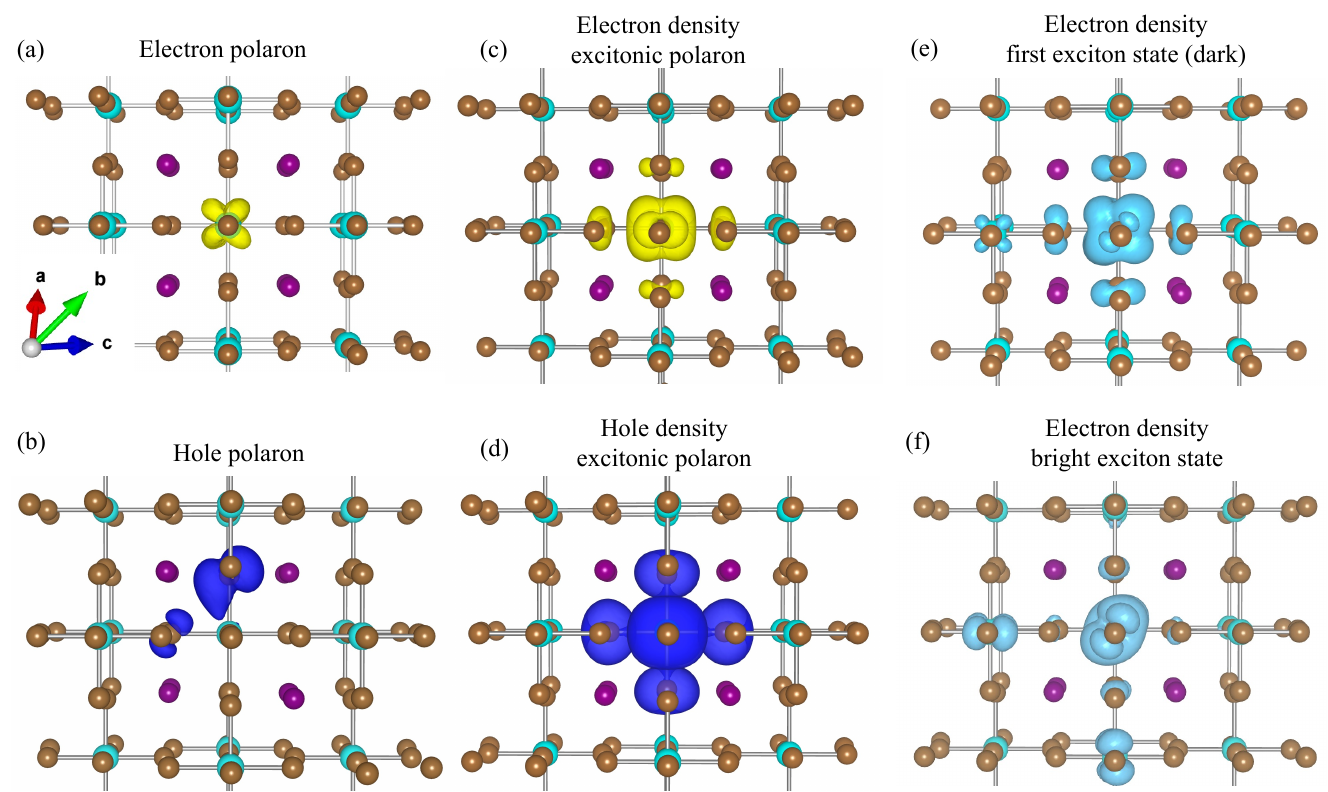}
\caption{
\label{fig:charge_density} 
% {\color{red}[Z: suggested changes: rotate figures to align octahedra with vertical axis; choose more interesting color scheme; change of labels; addition of tags for each panel; wfont sizes/types same as in caption; let's use the whole space with nice big panels and smaller spacing between them, see layout in black]}
% {\color{red}[Z: can you please change the labels of (e), (f) into "first exciton state (dark)", "bright exciton state"; also the labels look low-res, maybe better to make sure this is in vector PDF]}
Charge density isosurfaces of electron polaron, hole polaron, exciton polaron, and excitons in \ch{Cs2ZrBr6}.
The atomic color code is: Cs purple, Zr cyan, Br brown.
(a) Charge density isosurface of the electron polaron (yellow). 
(b) Charge density isosurface of the hole polaron (dark blue). 
(c) Electron density isosurface $n_{\rm e}({\bf r}_{\rm e})$ of the excitonic polaron (yellow).
(d) Hole density isosurface $n_{\rm h}({\bf r}_{\rm h})$ of the excitonic polaron (dark blue).
(e) Electron density isosurface of the lowest energy exciton (dark) of \ch{Cs2ZrBr6} in the ground-state undistorted structure (light blue).
(f) Electron density isosurface of the bright exciton corresponding to the first absorption peak of \ch{Cs2ZrBr6}, as shown in Fig.~S4(a), in the ground-state undistorted structure  (light blue).
In (e) and (f), the hole position ${\bf r}_{\rm h}$ is chosen to coincide with a Br atom.
}
\end{figure*}

\begin{figure*}[t]
\includegraphics[width=0.8\textwidth]{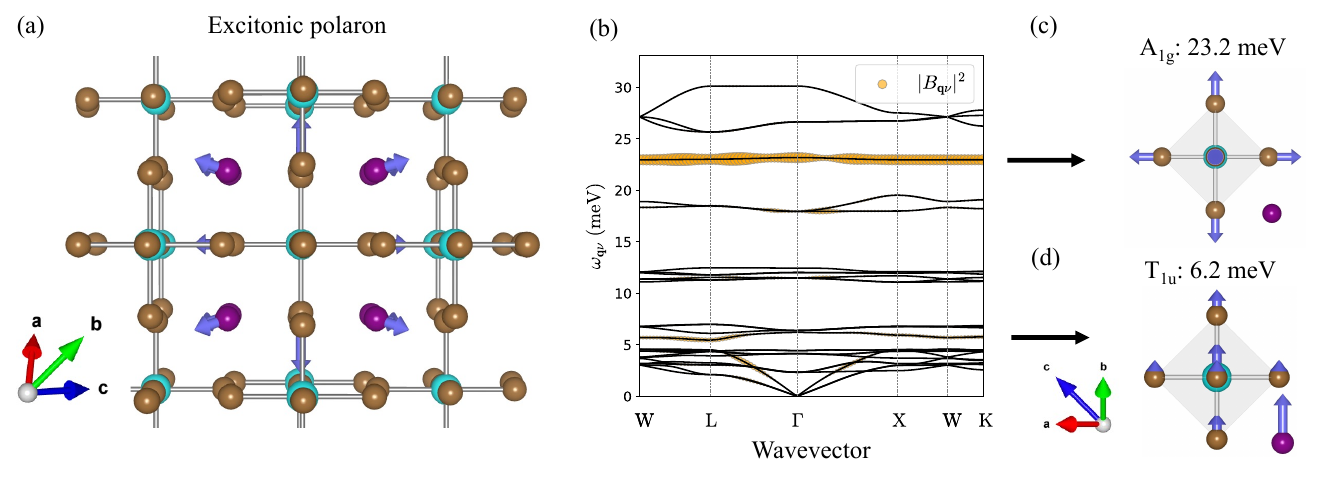}
\caption{
% {\color{red}[Z: can you please make the fonts a little better/same size everywhere and vector PDF? Also I would remove the rectangle legend in (d) which is redundant. In (e) and (f), let's use one decimal digit for consistency, 4.4 and 23.2. The displacementsin (e) and (f) are not clear, can we remove the smallish ones and delete atoms to show a single octahedron in each case.]} 
Atomic displacements accompanying the excitonic polaron of \ch{Cs2ZrBr6}. 
We only display significant displacements for clarity.
The atomic color code is: Cs purple, Zr cyan, Br brown.
(a) Atomic displacements (blue arrows) associated with the excitonic polaron, whose electron and hole charge densities are shown in Figs.~\ref{fig:charge_density}(c) and (d), respectively.
(b) Phonon contribution to the formation of the excitonic polaron. Black lines are the phonon dispersion relations, and the yellow discs are proportional to $|B_{\bmq\nu}|^2$.
(c) Eigendisplacements of the $A_{\rm 1g}$ mode at 23.2~meV at the zone center.
(d) Eigendisplacements of the $T_{\rm 1u}$ mode at 6.2~meV at the zone center.
In (c) and (d), only one octahedron and one Cs atom are displayed since all other octahedra and Cs atoms have the same displacement pattern.
\label{fig:phonon} 
} 
\end{figure*}

\begin{figure}[h]
\includegraphics[width=\columnwidth]{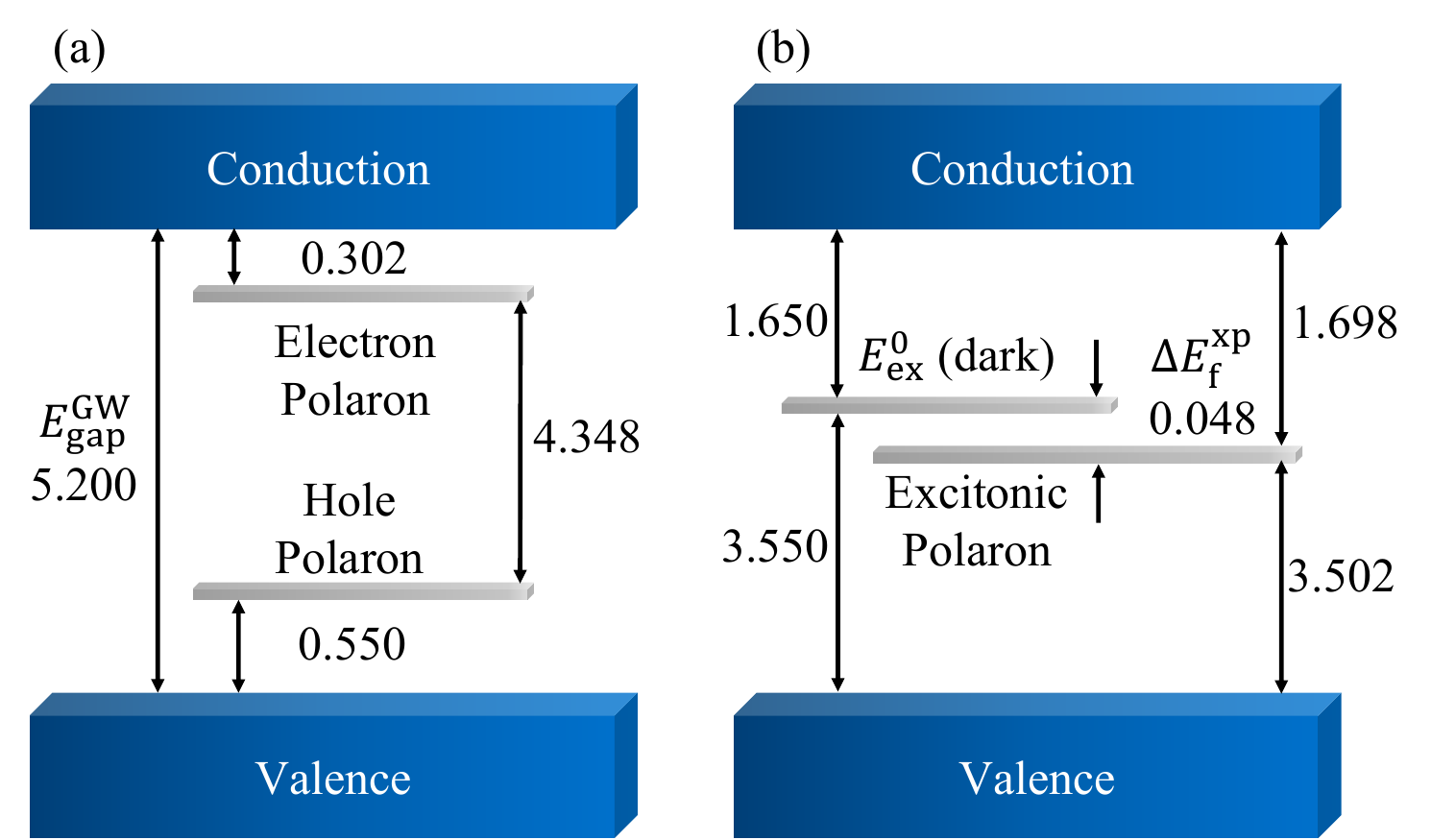}
\caption{
\label{fig:energies} {Energetics of polarons and excitonic polarons in \ch{Cs2ZrBr6}. {\color{black}All values in (a) and (b)} are in the unit of eV. (a) GW quasiparticle band gap and formation energies of the electron polaron and of the hole polaron. (b) Energetics of exciton and excitonic polaron. On the left-hand side of this panel we show the excitation energy of the lowest (dark) exciton state with respect to the ground state, and its binding energy as measured with respect to the GW quasiparticle band gap. On the right-hand side of this panel, we show the excitation energy of the excitonic polaron with respect to the ground state, the binding energy with respect to the quasiparticle gap, and the stabilization energy from the free exciton state to the excitonic polaron state.}}
\end{figure}

\clearpage


\section*{Supplementary material}

# Excitonic polarons and self-trapped excitons
# from first-principles exciton-phonon couplings

Zhenbang Dai,[1,2] Chao Lian,[1,2] Jon Lafuente-Bartolome,[1,2] and Feliciano Giustino[1,2,*]

[1]*Oden Institute for Computational Engineering and Sciences,*
*The University of Texas at Austin, Austin, Texas 78712, USA*
[2]*Department of Physics, The University of Texas at Austin, Austin, Texas 78712, USA*
(Dated: December 7, 2023)

## COMPUTATIONAL SETUP

To obtain the optimized structure, Kohn-Sham states and energies, and phonon eigenmodes and frequencies, we perform density functional theory (DFT) and density functional perturbation theory (DFPT) calculations using the QUANTUM ESPRESSO package [1, 2]. We employ the PBE generalized-gradient approximation to the exchange and correlation functional [3], norm-conserving pseudopotentials [4, 5], and a planewaves kinetic energy cutoff of 90 Ry. Spin-orbit coupling (SOC) is included since it was shown to have significant influence on the optical properties of $Cs_2ZrBr_6$ [6]. We use EPW [7, 8] and WANNIER90 [9] to compute the electron-phonon coupling matrix elements and polarons [10], and BERKELEYGW to perform GW/BSE calculations with finite exciton momentum [11–13]. We set the kinetic energy cutoff for the dielectric matrix to 15 Ry, and include 72 valence bands and 528 conduction bands. The BSE kernel is constructed using 18 valence bands and 6 conduction bands. Equations (7) and (8) are solved on a 4×4×4 Brillouin zone grid for the electron, phonon, and exciton momenta. This choice corresponds to an equivalent BvK supercell with 576 atoms. These choices are adequate to obtain a converged absorption spectrum at the band-edge, as shown in supplemental Fig. S4. The software VESTA is used for the visualization of the crystal structures, charge densities, and displacement patterns [14].

For the solution of Eqs. (7)-(8), the coefficients $A_{s\mathbf{Q}}$ are initialized such that $\Psi(\mathbf{r}_e, \mathbf{r}_h)$ is already sufficiently localized. To generate such a trial localized solution, we artificially enhance the exciton-phonon coupling strength by setting $g_{v'v\nu} = 0$ in Eq. (9). This choice does not affect the final self-consistent solution. Then, we evaluate $B_{\mathbf{Q}\nu}$ using Eq. (8), and we diagonalize the matrix in Eq. (7); we repeat this process until convergence is achieved.

We point out that, in our formalism, the excitonic polaron breaks translational invariance because nuclei are described within the adiabatic and classical approximation. In a fully many-body framework, the excitonic polaron wavefunction and the atomic displacements would be localized relative to each other, but the complete many-body state would be translationally invariant. Therefore, the object that we investigate here is a "pinned" excitonic polaron. We refer to Sec. VI of Ref. 15 for a detailed discussion of this point.

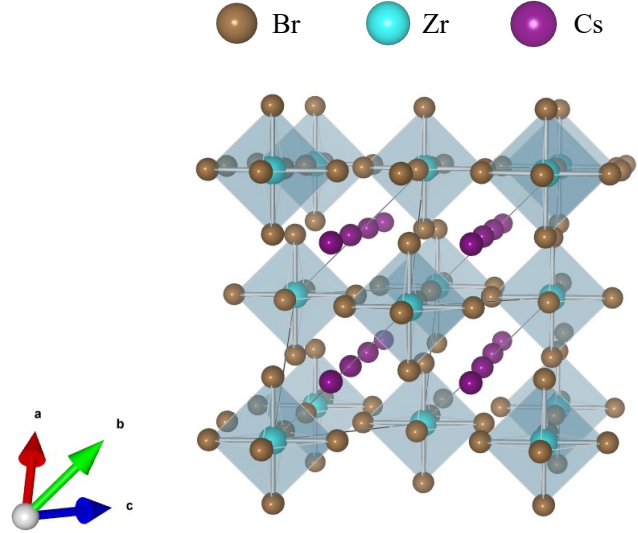

FIG. S1. Crystal structure of $Cs_2ZrBr_6$. The octahedra are "disconnected" from each other, a characteristic feature of vancancy-ordered double perovskites. The space group is $Fm\bar{3}m$ (225) [6].

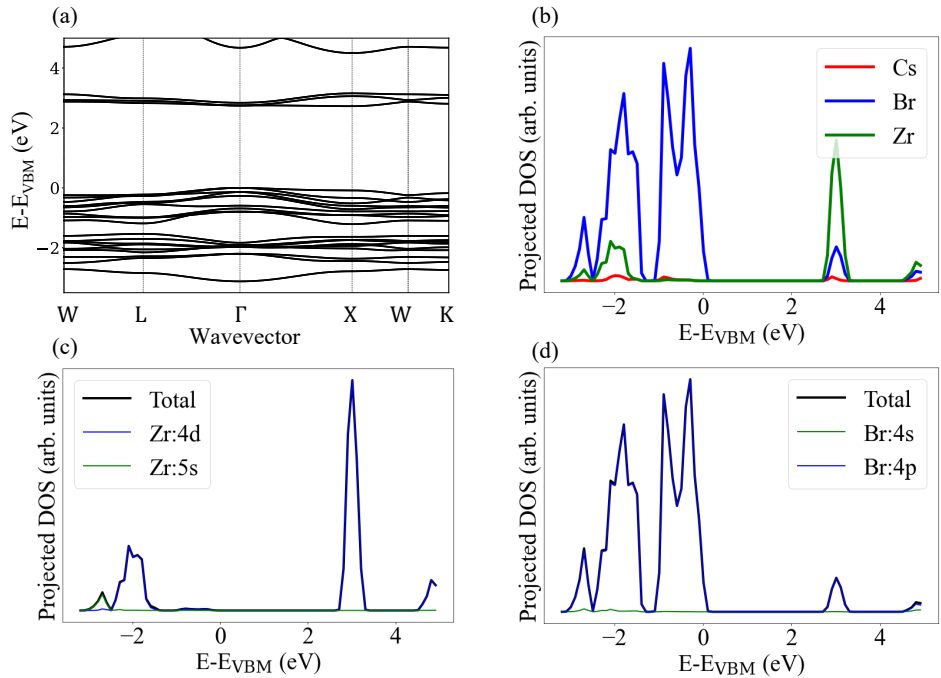

FIG. S2. (a) DFT band structure of $Cs_2ZrBr_6$. Since the the lowest three conduction bands are isolated from higher conduction bands, we include this isolated manifold in the solution of Bethe-Salpeter equations. (b) Decomposition of the density of states into contributions from Cs (red line), Br (blue line), and Zr (green line), respectively. (c) Decomposition of the density of states associated with Zr atoms into $4d$ (blue line) and $5s$ (green line) states. The black line is the total Zr density of states. (d) Decomposition of the density of states associated with Br atoms into $4p$ (blue line) and $4s$ (green line) states. The black line is the total Br density of states.

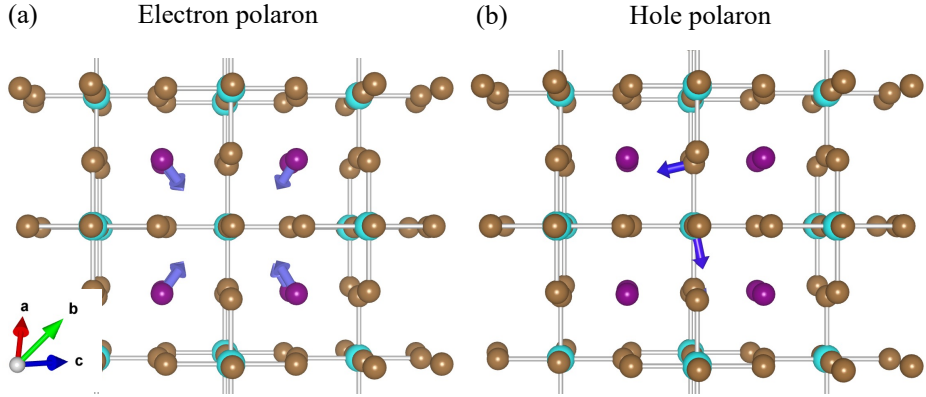

FIG. S3. Atomic displacements accompanying the electron polaron and the hole polaron of $Cs_2ZrBr_6$. We only display significant displacements for clarity. The atomic color code is: Cs purple, Zr cyan, Br brown. (a) Atomic displacements (blue arrows) associated with the electron polaron, whose charge density is shown in Fig. 1(a). (b) Atomic displacements (blue arrows) associated with the hole polaron, whose charge density is shown in Fig. 1(b).

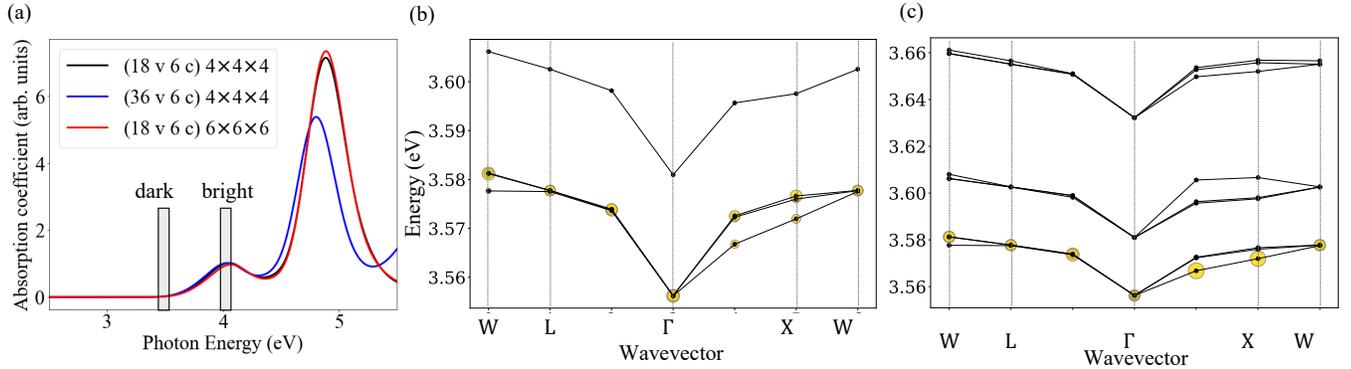

FIG. S4. (a) Sensitivity of the calculated optical absorption spectrum of $Cs_2ZrBr_6$ on the parameters used to solve the Bethe-Salpeter equations. We investigate the sensitivity of the spectrum to the number of bands included in the BSE equations (18 valence bands and 6 conduction bands, or 36 valence bands and 6 conduction bands), and Brillouin-zone sampling (4×4×4 and 6×6×6). When using 18+6 bands and 4×4×4 grid points, the absorption onset around 3.5 eV is accurately reproduced as compared to calculations where more bands and denser $\boldsymbol{k}$-grid are used; this is the energy region which contributes the most to the formation of the excitonic polaron, as shown in (b) and (c). The vertical bars represent the lowest exciton state (dark) and the first strong absorption peak (bright exciton). (b)-(c) Exciton band structure and contributions of individual exciton states to the excitonic polaron. Black lines are the band structure, yellow disks are proportional to the magnitude of the expansion coefficients $|A_{s\mathbf{Q}}|^2$. (b) The excitonic polaron equations are solved by including 4 exciton bands. (c) The excitonic polaron equations are solved by including 9 exciton bands. In both panels, the coefficients $|A_{s\mathbf{Q}}|^2$ are concentrated in the three lowest-energy bands, and are distributed nearly uniformly across the Brillouin zone.

* fgiustino@oden.utexas.edu


\end{document}